\documentclass[12pt,preprint]{aastex}
\begin{document}

\title{Dark Energy and the Hubble Constant}

\author{H. Arp}
\affil{Max-Planck-Institut f\"ur Astrophysik, Karl Schwarzschild-Str.1,
  Postfach 1317, D-85741 Garching, Germany}
 \email{arp@mpa-garching.mpg.de}

\begin{abstract}

Dark energy is inferred from a Hubble expansion which is slower at epochs
which are earlier than ours. But evidence reviewed here shows $H_0$ for nearby
galaxies is actually less than currently adopted and would instead
require {\it deceleration} to reach the current value.

Distances of Cepheid variables in galaxies in the Local Supercluster
have been measured by the Hubble Space Telescope and it is argued
here that they require a low value of $H_0$ along with redshifts which are at
least partly intrinsic. The intrinsic component is hypothesized to be
a result of the particle masses increasing with time.

The same considerations apply to Dark Matter. But with particle masses
growing with time, the condensation from plasmoid to proto galaxy not
only does away with the need for unseen ``dark matter'' but also
explains the intrinsic (non-velocity) redshifts of younger matter.

\end{abstract}


{\bf Introduction}

Recent analysis of supernovae data yields $H_0$ = 65 km/sec/Mpc up
to a redshift of z = .35 (Shafieloo 2007). If we take the currently
accepted value out to about 25 Mpc in our own neighborhood then 
$H_0$ = 72 km/sec/Mpc. (Freedman et al. 2001). This is interpreted as
an acceleration in an expanding universe from 65 to 72 in the time
from z = .35 to the present.

But if the Hubble constant appropriate to our nearby galaxies
is $H_0$ = 55 (not 72) we are left with a deceleration of -10 km/sec/Mpc
instead of an acceleration of + 7 km/sec/Mpc. As an average this would
indicate a contraction (Due to negative Dark Energy?) of $-1.5 {\pm 8.5}$
km/sec/Mpc - an imprecision suggestive of no meaningful evidence for
dark energy. (See Table 1).

\begin{figure}
\includegraphics[width=15.0cm]{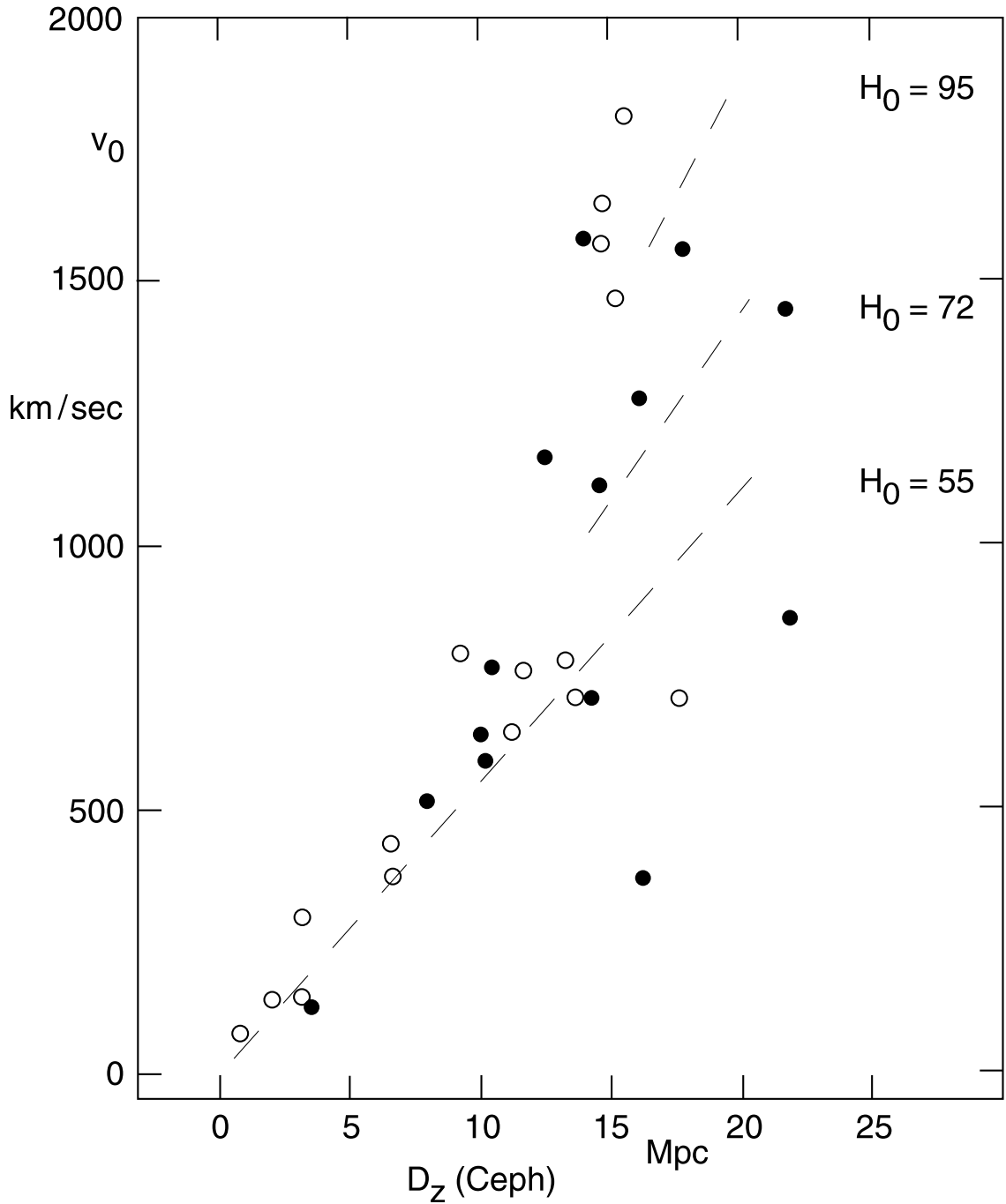}
\caption{The Cepheid distance, $D_z$(Ceph), is plotted against Local
Group centered redshift ($v_0$) for available galaxies. For low
redshift galaxies a very accurate fit to $H_0$ = 55 is
evident. Greater than 800 km/sec, however, excess redshifts appear
which are too large and too positive to be peculiar velocities. Filled
circles are Sb, open circles Sc.
\label{fig1}}
\end{figure}

\begin{table}[t]
\caption{Dark Energy\label{tab:galdist}}
\vspace{0.4cm}
\begin{center}
\begin{tabular}{ c c c l}
& & & \\
Region  &~~~~
$H_0$ &
$Dark Energy$ &
\\ \hline

z = .35 &~~~~
65 &~~~~ \\

Local SuperCluster &~~~~
72 &~~~~
+7 km/sec/Mpc& \\

Local SuperCluster &~~~~
55 &~~~~
-10 km/sec/Mpc& \\

& & &  \\ 
\end{tabular}
\end{center}
\end{table}

\section{What is $H_0$ Locally?}

In view of the contradictory results obtained from the analysis of the
HST data on local $H_0$, it is important to see if any dark energy
values can be reconciled.

The Cepheid distances are considered to be the most accurate distances 
available. The Hubble Space Telescope was used to measure light curves 
in order to obtain mean magnitudes and periods. The most illuminating
display of the data is in the redshift - distance diagram shown here
in Fig. 1. Clearly the nearest spirals with Cepheids define accurately 
the $H_0$ = 55 line. Greater than about 800 km/sec, however, the
deviations are well above the $H_0$ = 55 line.

The galaxies which fall above the line at higher redshifts 
are predominantly high luminosity class, high z spirals. The evidence
for intrinsic redshifts then lies in the fact that, for luminosity
class I galaxies, their redshifts are much too high to give reasonable
Hubble constants. (See Table 1 of Arp 2002.) 

Another way of judging the existence of intrinsic (non velocity)
redshifts is to examine the Virgo Cluster. It has been known for a
long time that the late type spirals appear to be members of this  
rich cluster. But it was argued by G. de Vaucouleurs among others that
these spirals, because of their higher redshift, were actually a
separate cluster at a greater distance behind, and only accidently
aligned with, the Virgo Cluster. It did not swing opinion when it was
shown that Sc galaxies in mixed groups and clusters exhibited
sytematically higher redshifts (Arp 1990, 1998). With the recent
Cepheid distances, however it is now possible to make a definitive
test for the Virgo spirals. They turn out to be 15 Mpc distant,
almost exactly at the center of the Virgo Cluster (Fig. 1). The excess
redshift of spiral galaxies which contain type I Cepheids then requires a
smaller $H_0$ which gives no evidence that would require Dark Energy.

\begin{figure}[ht]
\includegraphics[width=11.0cm]{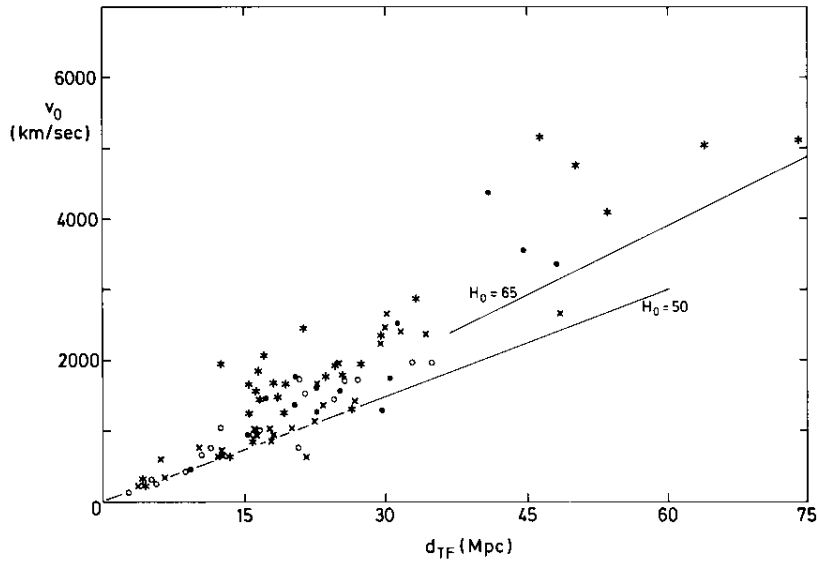}
\caption{The Tully-Fisher distance ($d_TF$) is plotted against Local
Group centered redshift ($v_0$) for Sc's. Asterisks denote Local Group
direction. For low redshift galaxies a very accurate fit is
evident. Data and Figure from Arp (1990).
\label{fig2}}
\end{figure}

\begin{figure}[ht]
\includegraphics[width=11.0cm]{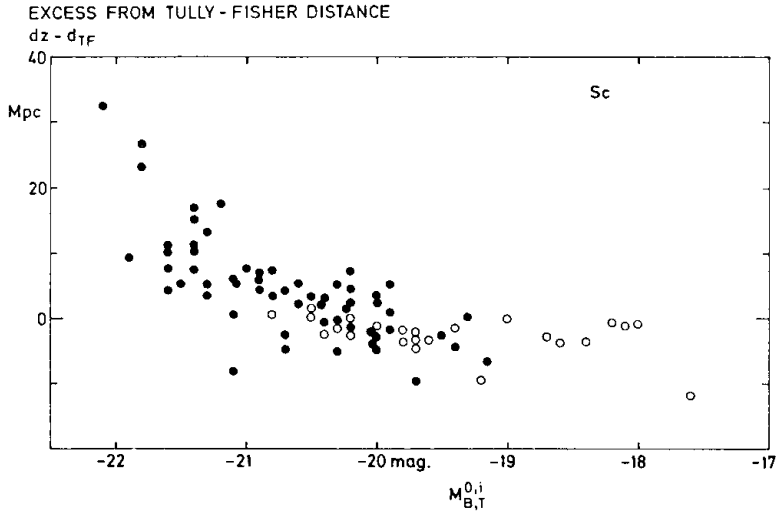}
\caption{The excess of redshift distance over Tully-Fisher distance
$(d_z - d_{TF})$ is plotted as a function of blue luminosity as
derived in Arp (1990). Open circles are Sc's with $v_0 \leq 1000$ km/sec
and demonstrate that redshift and Tully-Fisher distances agree well
for low redshift galaxies over a wide range in luminosity.
\label{fig3}}
\end{figure}

\subsection{Independent Checks on Redshift Distances}

Two of the strongest independent confirmations of excess redshifts are:
     
     1) Tully-Fisher distances from rotational velocity data. Fig. 2
        shows that Tully/Fisher within 15 Mpc agrees well with Cepheid  
        distances and low $H_0$ but that at greater distances the
        redshifts become excessive. Unless space is expanding faster
        in earlier times (opposite to that of Dark Energy) the
        redshifts must be intrinsic. 

     2) If we take the redshift of the brightest galaxy in Virgo -
        presumably the most massive in the cluster - we obtain: 

       $$ H_0 = 822/15.3 = 53.7 km/sec/Mpc $$  

It is also interesting to note that Sandage and Tamman regularly
obtained $H_0$ near 55. They stress, however the use of volume limited
samples. (Sandage, Tamman and Saha 1998). Ironically their volume was
thus defined by using redshift distances which thereby excluded the excess
(intrinsic) redshift galaxies. Hence they obtained the low but more correct
$H_0$ for galaxies near the age of our Milky Way. Note, however, their
$H_0$ = 62 in a later group paper (Sandage et al. 2006) where they
extend the redshift volume to about 20,000 km/sec. Naturally they now
include some of these intrinsic redshift galaxies which
raise the Hubble constant above the earlier, more correct, derived
values. We see from Fig.3 that the redshift distances for such
galaxies can give distances discrepant by up to 30 Mpc. 

Finally Morley Bell (2007) notes ``. . . the local Hubble constant is
found to be $H_0$ = 58 km/sec/Mpc when the intrinsic components are
removed.''

\section{The variable Mass theory}

At this point it would help to place the observational phenomenon of
intrinsic redshift into a theory which could connect it to the world
of accepted physics. We look first at the Einstein field equations,
the fundamental energy/momentum conservation statement that furnishes
the departure point for rigorous cosmological theories.  The key to
the general solution of the equations based on the Hoyle/Narlikar
Machian gravitation theory is elementary particle masses varying as m
= m(t) Narlikar (1977). For a constant mass approximation the theory
reduces to the standard Einstein theory. However, the input from
Mach's principle suggests that the inertia of a newly created particle 
starts off as zero and grows with its age as it begins to get
contributions from more and more remote matter in the universe. The
typical wavelengths emitted by a particle (such as the electron in a
hydrogen atom) would reduce as its mass grows. Thus newly created
matter would exhibit high intrinsic redshift. In such a framework, the
intrinsic redshift of all galaxies created at the same time as our own
will always give a perfect Hubble relation because the look back time
to a distant galaxy will always reveal it at a younger age when its
intrinsic redshift was exactly that predicted by the Hubble law,  
cz = d x Ho. Younger galaxies would have intrinsic redshifts.

\subsection{The Indeterminancy of Dark Energy}

We observe redshifted supernovae at earlier stages in their evolution
due to look-back time. If they are young enough to have appreciable
intrinsic redshift then we would calculate a higher $H_0$ at the epoch
of the supernova and we would have to slow down the supposed cosmic
expansion to match our current $H_0$. This would be opposite to the
dark energy being widely discussed at present.

Of course younger galaxies that are evolving the mass of their
elementary particles would shine with a somewhat weaker light
than nearby standard supernovae so we would tend to derive a greater
than true distance for them. The resulting Hubble constant could be
smaller than for our local neighborhood and we might conclude that the
universe is now expanding faster than in the past. 

The two effects would tend to vie with each other for dominance at any
given time. And since both are very difficult to calculate the true
expansion velocity, if indeed there is any expansion, would be
indeterminant regardless of how accurately our local $H_0$ were to be
agreed on. Simon White presciently remarked anent Dark Energy:
''Unfortunately, the progenitors of higher redshift supernovae formed
and exploded in younger galaxies than their lower redshift
counterparts, and this could plausibly cause small redshift-dependent
shifts in the properties of the supernovae or of their immediate
environments. Undetected shifts of this kind could confuse the search
for the Dark Energy signal and limit the precision with which it can
be measured'' (White, S. 2007 arXiv: 0704.229).             

\subsection{Suggestions of Local Expansion}

Recently Chernin et al(2007) have interpreted long standing observations
of groups of galaxies in terms of cells of dark matter expansion. The
problem here is that analyses of these groups have shown repeatedly
that the smaller companions around the dominant group galaxy are of
systematically higher redshift.(E.g. 22 out of 22 major companions to
M 31and M 81 have higher redshifts). (Arp 1994;1998a;1998b) If the groups
were expanding one would expect as many approaching as receding
galaxies as we obeserve them. In fact the observations in many groups
and clusters demonstrate the reality of the intrinsic redshift, the
same redshift that requires the low $H_0$ in our Local Group galaxies
and which, as the preceding discussion argues, points to dark energy as
not having been detected.

\subsection{Dark Matter and Formation of Galaxies}

The evidence for intrinsic redshifts discussed in the determination of
local $H_0$ is pertinent to the subject of dark matter in galaxy
formation.  With particle masses growing with time as $t^2$ 
(Narlikar and Arp 1993), the condensation from plasmoid to proto
galaxy in the early stages is not only accomplished without the need
for unseen ``dark matter'' but also explains the intrinsic redshifts
of younger matter. For example the long debated association of high redshift
quasars with active low redshift galaxies then leads to a continuous
evolution from quasars through active galaxies to older galaxies on
the Hubble relation in a continuously creating Universe. (See
arXiv.0711.2607 for discussion of Quasars and the Hubble relation.)

\noindent{\bf References}
\medskip 

\noindent Arp, H. 1990, Ap\&SS 167, 183
\medskip

\noindent Arp, H. 1994, 430, 74
\medskip

\noindent Arp, H. 1998a, Seeing Red, Apeiron, Montreal
\medskip

\noindent Arp, H. 1998b, ApJ 496, 661

\noindent Arp, H. 2002, ApJ 571, 615  
\medskip

\noindent Bell, M. 2007, ApJ 667, L129
\medskip

\noindent Freedman,W., Madore, B., Gibson, B. et al. 2001, ApJ 553, 47
\medskip

\noindent Narlikar J. 1977, Ann. Phys. 107, 325
\medskip

\noindent Narlikar J. and Das, P. 1980, ApJ 240, 401
\medskip

\noindent Narlikar J., Arp H. 1993 ApJ 405, 51
\medskip

\noindent Sandage A., Tamman, G., Saha, A. 1998 Phys. Rep. 307, 1 
\medskip
 
\noindent Sandage, A. et al. 2006, ApJ 653, 843
\medskip

\noindent Shafieloo. A. 2007, Mon. Not. Roy. Soc. 380(4), 1573
\medskip

\noindent White, S. 2007, astro-ph, arXiv:0704.229
\medskip

\end{document}